\documentclass[11pt]{article}
\usepackage{amssymb, latexsym, amsbsy, amsfonts}
\newtheorem{theorem}{Theorem}[section]

\newtheorem{lemma}[theorem]{Lemma}

\newtheorem{rem}{Remark}
\newcommand{\remark}{\begin{rem} \normalfont}\def\endremark{\end{rem}}

\newcommand{\cov}{\mathop{\mathrm{cov}}\nolimits}

\newcommand{\p}{\mathop{\mathrm{{\cal P}}}\nolimits}
\newenvironment{pf}{\noindent \textit{Proof}: \ignorespaces}{\hfill\ensuremath
{\Box}\par}

\makeatletter
\@addtoreset{equation}{section}
\makeatother

\newcommand {\R} {\ensuremath{\mathbb{R}}}

\newcommand {\n}  {\ensuremath {\mathsf{N}}}

\newcommand {\Z}  {\ensuremath {\mathcal{Z}}}

\newcommand {\N}  {\ensuremath {\mathcal{N}}}
\newcommand {\G}  {\ensuremath {\mathcal{G}}}
\newcommand {\g}  {\ensuremath {\mathsf{G}}}
\newcommand {\pp}  {\ensuremath {\mathsf{P}}}

\newcommand {\rh} {\ensuremath{\varrho}}

\begin{document} 
%\title
\begin{titlepage}
  \begin{center}
    {\Large \bf Infrared regular representation of the three dimensional massless Nelson model}
      
\end{center}
\vspace{0.2cm}
\begin{center}
  
    \small J\'ozsef L\H{o}rinczi$^1$, Robert A. Minlos$^2$ and Herbert Spohn$^1$\\[0.1cm]
\end{center}
\vspace{1.5cm}
%\makebox{
%\begin{minipage}
    {\it \small $^1$Zentrum Mathematik, Technische Universit\"at M\"unchen} \\
    {\it \small 80290 M\"unchen, Germany} \\
    {\small lorinczi@mathematik.tu-muenchen.de}\\
    {\small spohn@mathematik.tu-muenchen.de}\\ [0.5cm]
%{\small and} \\
%\small Robert A. Minlos\\[0.1cm]
    {\it \small $^2$Dobrushin Mathematics Laboratory} \\
    {\it \small Institute for Information Transmission Problems} \\
    {\it \small Bolshoy Karetny per. 19, 101447 Moscow, Russia }\\
    {\small minl@iitp.ru}\\ [1cm]
%}
%\end{minipage}
\vspace{1.5cm}    
\date{}
\begin{abstract} 
We prove that in the Euclidean representation of the three dimensional massless Nelson model 
the $t=0$ projection of the interacting measure is absolutely continuous with respect to a
Gaussian measure with suitably adjusted mean. We also determine the Hamiltonian in the Fock
space over this Gaussian measure space.
\end{abstract}
\end{titlepage}
%\maketitle
%\baselineskip 0.8 cm
\vspace{1cm}
\section{Introduction}
The Nelson model \cite{Nel} of a spinless electron coupled to a scalar massless Bose field is 
infrared divergent in 3 space dimensions. In its algebraic version the ground state 
representation of the Nelson Hamiltonian is not unitarily equivalent to the Fock representation 
\cite{F,AHH}. From the point of view of functional integrals, which is the one taken here, the 
$t=0$ projection of the interacting measure is singular with respect to the $t=0$ projection of 
the free measure \cite{LMS}. Such a result leaves open the possibility of an explicit 
representation of the $t=0$ functional measure. In this letter we will settle the issue. In the
last section we will explain the connection to the non-Fock representation. 

The infrared divergence in Nelson's model can be seen already on the level of the classical 
action. The Euclidean action for the electron is
\begin{equation}
S_{\mathrm{p}}(\{q_t\}) = \int_{-\infty}^\infty \left(\frac{1}{2} \dot q_t^2 + V(q_t) \right) dt.
\end{equation}
Here $t \mapsto q_t \in \R^3$ is a path of the electron. The electron's mass is set equal to one
only for convenience. $V: \R^3 \to \R$ is a confining potential, which for technical reasons we
require to be bounded from below, continuous and have the asymptotics $V(x) = C |x|^{2s} + 
o(|x|^{2s})$ for large $|x|$ with some constant $C > 0$ and exponent $s > 1$. The massless field
$\xi_t: \R^3 \to \R$ has the action 
\begin{equation}
S_{\mathrm{f}}(\{\xi_t\}) = \int_{-\infty}^\infty\int_{\R^3} \frac{1}{2} 
\left(\partial_t\xi_t(x)^2 + \nabla \xi_t(x)^2 \right) dx dt. 
\end{equation}
Finally, the electron is coupled locally and translation invariant as given by
\begin{equation}
S_{\mathrm{int}}(\{q_t, \xi_t\}) = \int_{-\infty}^\infty \int_{\R^3} \rh(x-q_t) \xi_t(x) dx dt. 
%\equiv \int_\R (\xi_t * \rh)(q_t) dt.
\end{equation}
Here $\varrho$ is a form factor, assumed to be a sufficiently smooth function and of rapid
decrease, normalized such that $\int_{\R^3} \varrho(x) dx = e$, where $e$ is the charge of the 
particle. The case of interest is $e \neq 0$, since $e=0$ implies $\hat\rh(0) = 0$ 
($\hat\rh$ the Fourier transform of $\rh$) and therefore regularizes the interaction in the 
infrared region. At a later stage we will need that $|e| \leq e^*$ with some sufficiently small 
$e^*$.

The infrared divergence can be deduced from the solution of the classical field equation which 
minimizes energy. An easy computation leads to the minimizer $q_t  = q_{\rm{\tiny{min}}}$, where
$q_{\rm{\tiny{min}}}$ corresponds to an absolute minimum of the potential, i.e. $V(x) \geq
V(q_{\rm{\tiny{min}}})$, for all $x \in \R^3$, and $\xi_t = \xi_{\rm{\tiny{min}}}$ given by
\begin{equation}
\xi_{\rm{\tiny{min}}} (x) = \Delta^{-1} \varrho (x - q_{\rm{\tiny{min}}}).
\end{equation}
In three dimensions $\xi_{\rm{\tiny{min}}} (x) \cong -e/(4\pi |x|)$ for large $|x|$. On the other
hand, for the free measure $\rh = 0$ and $\xi_{\rm{\tiny{min}}} = 0$. If the total action $S =
S_{\mathrm{p}} + S_{\mathrm{f}} + S_{\mathrm{int}}$ is used for constructing the Euclidean path 
measure (formally like $\exp(-S)$), then basically one has to compare two Gaussian fields of the 
same covariance, one with mean $\Delta^{-1} \varrho(x)$ and the other with zero mean. Because of 
slow decay of the Coulomb potential in 3 dimensions, these two measures cannot have a density with 
respect to each other. 

Our arguments also offer a hint of how to improve on this situation. We reshuffle the field 
action by adding a term dependent on a function $u$ to be chosen below:
\begin{equation}
S_{\mathrm{f}}^u (\{\xi_t\}) = \int_{-\infty}^\infty\int_{\R^3} \left( \frac{1}{2}  
\left (\partial_t\xi_t(x)^2 + (\nabla \xi_t(x))^2 \right) + \xi_t(x)u(x) \right) dx dt. 
\end{equation}
For convenience, we represent this auxiliary function %$u$ 
as the convolution
\begin{equation}
u(x) = (\rh * h)(x) \equiv \int_{\R^3} e^{i k\cdot x}\hat\rh (k)\hat h(k) dk 
\label{hrh}
\end{equation}
where $\hat h$ denotes the Fourier transform of $h$, of which we require that 
\begin{enumerate}
\item
\mbox{$\hat h$} is a real-valued, even, bounded, and sufficiently smooth function;
\item
\mbox{$\hat h (0) = 1$}.
\end{enumerate}
The first condition is more for simplicity than necessity, the second is essential. The 
density $\exp(-S_{\mathrm{f}}^u (\{\xi_t\}))$ provides the free Gaussian measure. Keeping 
the electron's action and the total action unchanged, we obtain a reshuffled interaction 
of the form
\begin{equation}
S_{\mathrm{int}}^u (\{q_t, \xi_t\}) 
= \int_{-\infty}^\infty \int_{\R^3} \left (\rh(x-q_t) - (\rh * h)(x) \right) \xi_t(x) dx dt. 
\label{su}
\end{equation}

The function $u$ is chosen in such a way that the free Gaussian's mean asymptotically 
agrees with the solution of minimal energy of the classical field equation. This then makes 
up for the possibility of the $t=0$ interacting measure to be absolutely continuous with 
respect to the $t=0$ projection of the above Gaussian measure. In the remainder of the
letter this argument will be made sound.

\section{Infrared regular representation}
First we translate the actions to well defined measures on path space. $S_{\mathrm{p}}$ is 
associated with the stationary $P(\phi)_1$-process $q_t$ on $C(\R,\R^3)$ defined by the 
stochastic differential equation
\begin{equation}
dq_t = (\nabla \log \psi_0) dt + dB_t
\end{equation}
($dB_t$ denotes Brownian motion), with invariant measure $d\n^0 = \psi_0^2(x) dx$. Here 
$\psi_0$ is the ground state of the Schr\"odinger operator $(-1/2)\Delta + V(q)$ generating 
the $P(\phi)_1$-process. The path measure of this process will be denoted by $\N^0$. The 
paths $Q = \{q_t: t \in \R\}$ are $\N^0$-almost surely continuous and $q_t$ is a 
time-reversible Markov process. Moreover, for our class of potentials $V$ the semigroup of 
this process is intrinsically ultracontractive \cite{LM}.

The field $\{ \xi_t (f): \; f \in {\cal S} (\R^3), t \in \R\}$ with $f \in {\cal S} (\R^3)$, i.e. 
the Schwartz space over $\R^3$, is in its turn described by the Gaussian measure $\G$ on 
$C(\R,\mathcal{S}'(\R^3))$ resulting from the (modified) Euclidean action $S_{\mathrm{f}}^u$.
This Gaussian measure has mean 
\begin{equation}
\mathbb{E}_{\G} [\xi_t(f)] = - \int_{\mathbb{R}^3} \frac{\hat\rh (k) \hat f(k) \hat h(k)}{|k|^2} dk 
\equiv \int_{\R^3} f(x) \gamma(x) dx
\label{shim}
\end{equation}
(i.e., represented in position space in terms of the function $\gamma$), and covariance 
\begin{equation}
\cov_{\G} \left (\xi_s(f_1), \xi_t (f_2) \right ) = \int_{\R^3} \frac{\hat{f}_1 (k) 
\hat{f}_2^*(k)}{2|k|} e^{-|k||t-s|} dk.
\label{gc}
\end{equation}
We denote the $t=0$ stationary distribution of this Markov process by $\g$; it is itself a 
Gaussian measure with mean (\ref{shim}) and covariance (\ref{gc}) taken with $s=t$. For easing
the argument below, we introduce the centred field
\begin{equation}
\eta_t (f) = \xi_t (f) - \mathbb{E}_{\G} [\xi_t(f)], \;\;\; f \in {\cal S}(\R^3);
\end{equation}
as an auxiliary variable; this obviously has then a Gaussian distribution with zero mean and the 
same covariance as (\ref{gc}) above. Note that $u$ introduces a shift in the mean of the unmodified 
Gaussian measure $\G_0$ on the same space (which with $h \equiv 0$ would be 0). We denote by $\g_0$ 
the $t=0$ distribution of $\G_0$. This shift induces the unitary map 
\begin{equation}
{\cal U}: L^2({\cal S}'(\R^3),d\g) \to L^2({\cal S}'(\R^3),d\g_0), \;\;\; 
({\cal U}F)(f) = F(f - \gamma), 
\label{ufi}
\end{equation}
with $F: {\cal S}'(\R^3) \to \R$, $f \in {\cal S}'(\R^3)$ and $\gamma$ given by (\ref{shim}),
which we introduce here for later use. 

The non-interacting joint particle-field process is set up on $C(\R,\R^3 \times \mathcal{S}'
(\R^3))$ and described by the path measure $\p^0 = \N^0 \times \G$ with $t=0$ distribution 
$\pp^0 = \n^0 \times \g$. The interacting system can be described by taking $\p^0$ as reference 
process and modifying it with the density given by 
\begin{equation}
\frac{d\p_T}{d\p^0}(q,\xi)  = \frac{1}{Z_T} \exp \left( -\int_{-T}^T S_{\mathrm{int}}^u
(\{q_t, \xi_t\}) dt \right) 
\label{fpp}
\end{equation}
Here
\begin{equation}
Z_T  = \int \exp\left(-\int_{-T}^T  S_{\mathrm{int}}^u(\{q_t, \xi_t\})  dt \right) d\p^0 
% = \int \exp\left(\int_{-T}^T \int_{-T}^T W(q_s-q_t, s-t) ds dt \right) d\N^0
\label{z}
\end{equation}
is the normalizing partition function. To have a more explicit formula instead of (\ref{z}) 
we first make the Gaussian part of the integration to obtain
\begin{eqnarray*}
Z_T 
& = &
\mathbb{E}_{\N^0_T} \left[ \mathbb{E}_{\G} \left[ \exp\left(-\int_{-T}^T \int_{\R^3} \hat \xi_t(k)
\hat\rh(k) (e^{i k\cdot q_t} - \hat h(k)) dk dt \right) \right] \right] \\ 
& = &
\mathbb{E}_{\N^0_T}  [ Z_T^Q ] 
\end{eqnarray*}
where
\begin{eqnarray}
\lefteqn{
Z_T^Q = 
\int \exp\left(-\int_{-T}^T \int_{\R^3} \hat \eta_t(k)
\hat\rh(k) (e^{i k\cdot q_t} - \hat h(k)) dk dt \right) \times}  \\ &&
\hspace{2.5cm}  \times \exp\left(\int_{-T}^T \int_{\R^3} \hat \gamma (k) 
\hat\rh(k) (e^{i k\cdot q_t} - \hat h(k)) dk dt \right) d\G_0 \nonumber \\ && 
= \exp \left(-\frac{1}{2} \int_{\R^3} \frac{|\hat \rh (k)|^2 \hat h(k)}{|k|} dk
\int_{-\infty}^\infty ds \int_{-T}^T dt e^{-|k||t-s|} (e^{i k \cdot q_t} - \hat h(k)) \right) 
\times  \nonumber \\ &&
\hspace{0.3cm} 
\times \exp \left( \frac{1}{4} \int_{-T}^T\int_{-T}^T \frac{|\hat\rh(k)|^2}{|k|} dk 
(e^{i k\cdot q_s} - \hat h(k))(e^{-i k\cdot q_t} - \hat h(k)) e^{-|k||t-s|} dtds \right).
\nonumber
\end{eqnarray}
Notice that in the above expression we passed from integration with respect to $\G$ to
integration with respect to $\G_0$. From this we get that
\begin{equation}
Z_T^Q = C_T \; {\cal Z}_T^Q
\end{equation}
with $C_T > 0$ independent of $Q$ and
\begin{eqnarray}
\lefteqn{
\Z_T^Q = \exp \left( -\int_{-T}^T\int_{-T}^T W(q_s-q_t,s-t) dtds \right) \times} 
\label{ww} \\ &&
\exp \left( -\int_{-T}^T dt \int_{|s| > T} ds \int_{\R^3} W(q_t-q,t-s) h(q) dq \right)  
\times                                                                       \nonumber \\ &&
\times \exp \left( -\int_{-T}^T ds \int_{|t| > T} dt \int_{\R^3} W(q_s-q,t-s) h(q) dq \right).
\nonumber
\end{eqnarray}
Here
\begin{equation}
W(q,t) = -\frac{1}{4} \int_{\R^3} \frac{|\hat\rh(k)|^2}{|k|} e^{-|k||t|} \;\cos (k \cdot q)\;dk
\label{w}
\end{equation}
This implies that the marginal distribution $\N_T$ of ${\cal P}_T$ for the particle paths
satisfies
\begin{equation}
\frac{d\N_T}{d\N^0} = \frac{\Z^Q_T}{\Z_T}.
\end{equation}
with $\Z_T = \int \Z_T^Q d\N^0$. The $t=0$ single time distribution of $\N_T$ will be 
denoted by $\n_T$. 
\begin{lemma}
There is some $e^* > 0$ such that for all $|e| \leq e^*$ the weak local limit 
$\lim_{T\to\infty} \N_T$ $ = \N$ exists and is a Gibbs probability measure on $C(\R,\R^3)$.
Moreover, $\N$ does not depend on $h$, i.e. it coincides with the limiting measure constructed 
for $h \equiv 0$. 
\end{lemma}
\begin{pf}
By using cluster expansion, the Gibbs measure for $h \equiv 0$ can be shown to exist for $W$ at 
sufficiently weak couplings $|e| \leq e^*$. Here we have an interaction which differs from the 
cases studied in \cite{LM} by the last two factors in (\ref{ww}). However, they appear as extra 
energies coming from interactions between pieces of paths lying inside $[-T,T]$ with constant 
boundary conditions in $\R \smallsetminus [-T,T]$ weighted by $h$, to which our previous arguments 
extend directly. 
\end{pf}

For any fixed path $ Q = \{q_t: t \in \R\} \in C(\R,\R^3)$ consider now the conditional 
distribution $\p^{Q}_T = \p_T(\;\cdot\;| \{q_t\} = Q)$ as a probability measure on $C(\R,\R^3 
\times {\cal S}'(\R^3))$. Since the interaction is linear in $\xi_t$, $\p^{Q}_T$ is itself a 
Gaussian measure with covariance
(\ref{gc}) and mean
\begin{equation}
\mathbb{E}_{\p^Q_T} [\xi_t(f)] = \int_{\R^3} \hat f(k) \hat g^{Q,t}_T (k) dk
\end{equation}
where 
\begin{eqnarray}
\hat g^{Q,t}_T (k) 
& = &
\hat\gamma(k) - \frac{\hat\rh(k)}{2|k|} \int_{-T}^T e^{-|k||t-\tau|} \left( e^{i k\cdot q_\tau} 
-\hat h(k) \right) d\tau \label{g} \\
& = &
- \frac{\hat\rh (k)}{2|k|} \int_{-T}^T e^{-|k||t-\tau|} e^{i k\cdot q_\tau} d\tau -
\frac{\hat\rh (k) \hat h(k)}{2|k|} \int_{|\tau| > T} e^{-|k||t-\tau|} d\tau. \nonumber 
\end{eqnarray}
\begin{lemma}
There is some $e^* > 0$ such that for all $|e| \leq e^*$ the weak local limit 
$\lim_{T\to\infty}\p_T$ $= \p$ exists and is a probability measure on $C(\R,\R^3 \times 
{\cal S}'(\R^3))$. Moreover, $\p$ coincides with the path measure for $h \equiv 0$. 
\end{lemma}
\begin{pf}
By using (\ref{g}) it is immediate that $\lim_{T\to\infty} \hat g^{Q,t}_T = \hat g^{Q,t}$ exists 
and coincides with the mean for $\p^Q$ with $h \equiv 0$. Thus we have also that $\p^Q_T \to \p^Q$ 
exists. Moreover, since $\p_T(E \times F) = \int_F \p_T^{ Q} (E) d\N_T$, for all $T > 0$, where 
$E$ and $F$ are appropriate measurable sets, we also have the claim of the lemma. Hence the joint 
particle-field path measure does not depend on $h$, i.e. on how the reshuffling of the free field 
and interaction energies is performed. 
\end{pf}
$\p$ and $\p^0$ are Markov processes whose $t = 0$ distributions we denote below by $\pp$ resp. 
$\pp^0$.

In the modified function space representation Nelson's Hamiltonian is $H$ on the Hilbert space 
${\cal H}^0 = L^2(\R^3 \times {\cal S}'(\R^3),d\pp^0)$ acting like
\begin{equation}
(F, e^{-tH} G)_{{\cal H}^0} = \mathbb{E}_{\p^0} \left[ F(q_0,\xi_0) G(q_t,\xi_t) \exp(-\int_0^t 
{\cal L}^u_{\mathrm{int}} (\{q_s, \xi_s\}) ds) \right], \;\;\; t > 0
\label{H}
\end{equation}
with suitable $F$ and $G$, and where ${\cal L}^u_{\mathrm{int}} (\{q_t, \xi_t\}) = \int_{\R^3} 
(\rh(x-q_t) - (\rh * h)(x)) \xi_t(x) dx$ is the Lagrangian corresponding to (\ref{su}). The 
bottom of the spectrum of $H$ will be denoted by $E_0$. We define the Euclidean Hamiltonian 
$H_{\rm{\tiny{euc}}}$ acting on the Hilbert space ${\cal H} = L^2(\R^3 \times {\cal S}'(\R^3),
d\pp)$ as the self-adjoint operator generating the symmetric contracting semigroup $T_t$ given by
\begin{equation}
(F, T_t G)_{\cal H} = \mathbb{E}_{\p}[F(q_0,\xi_0) G(\xi_t,q_t)], \;\;\; t > 0.
\label{euc}
\end{equation}
\begin{theorem}
Suppose $|e| \leq e^*$ with some $e^* > 0$. Then $H$ has a unique strictly positive ground 
state $\Psi_0$ corresponding to the lowest eigenvalue $E_0$, and $H - E_0$ is unitarily 
equivalent with $H_{\rm{\tiny{euc}}}$. 
\label{t}
\end{theorem}
\begin{pf}
From \cite{LMS} we know that the existence of a strictly positive ground state of $H$ implies 
that $H - E_0$ and $H_{\rm{\tiny{euc}}}$ are unitarily equivalent. Moreover, by Theorem 4.1 of 
the same reference we also know that for proving existence of the ground state 
%can be expressed in terms of the density $d\pp/d\pp^0$, whenever it exists. Thus 
it suffices to show that
\begin{equation}
\liminf_{T\to\infty} \int \left( \frac{d\pp_T}{d\pp^0} \right)^{1/2} d\pp^0 \;>\; 0.
\label{sufi}
\end{equation}
We have
\begin{eqnarray*}
\lefteqn{
\frac{d\pp^Q_T}{d\g} (\xi) = 
\exp \left( \int_{\R^3} (\hat\xi(k)- \hat\gamma(k))\hat\rh(k) dk 
\int_{-T}^T (e^{ik\cdot q_t} - \hat h(k)) e^{-|k||t|} dt \right) \times } \\ && 
\exp \left(- \frac{1}{4} \int_{\R^3} dk \int_{-T}^T \int_{-T}^T \frac{|\hat\rh (k)|^2}{|k|}
(e^{ik\cdot q_t}-\hat h(k))(e^{-ik\cdot q_s}-\hat h(k)) e^{-|k|(|t|+|s|)} dt ds \right).
\end{eqnarray*}
Moreover,
\begin{equation}
\frac{d\pp_T}{d\pp^0} (q,\xi) = 
\mathbb{E}_{\N_T} \left[ \frac{d\pp^Q_T}{d\g} (\xi) | q_0 = q \right] \frac{d\n_T}{d\n^0}(q).
\end{equation}
By using that for some $c > 0$
\begin{equation}
\frac{1}{c} \; \leq \; \frac{d\n_T}{d\n^0} \; \leq \; c
\end{equation}
obtained similarly as in \cite{LM}, in combination with Jensen's inequality we obtain 
\begin{eqnarray}
\lefteqn{
\int \left( \frac{d\pp_T}{d\pp^0} (q,\xi) \right)^{1/2} d\pp^0(q) \label{expo} } \\ &&
= \int \mathbb{E}_{\N_T} \left[ \left( \frac{d\pp^Q}{d\g} \right)^{1/2} | q=q_0 \right]
\left( \frac{d\n_T}{d\n^0} (q) \right)^{1/2} d\pp^0 \nonumber\\ && 
\geq \frac{1}{\sqrt c} \int \exp\left( \mathbb{E}_{\N_T}\left[ \frac{1}{2} \int_{\R^3} 
(\hat\xi(k) -\hat\gamma(k)) \hat\rh(k) \int_{-T}^T e^{-|k||t|}(e^{ik\cdot q_t} - 
\hat h(k)) dt dk|q=q_0 \right] \right) \nonumber \\ && 
\hspace{0.4cm} \times \exp \left( -\frac{1}{8} \int_{\R^3} dk \int_{-T}^T\int_{-T}^T 
\frac{|\hat\rh(k)|^2}{|k|} m_T(k;s,t) e^{-|k|(|t|+|s|)} dt ds \right) d\pp^0 \nonumber \\ &&
\geq \frac{1}{\sqrt c} \exp \left( -\frac{c}{8} \int_{\R^3} dk \int_{-T}^T\int_{-T}^T 
\frac{|\hat\rh(k)|^2}{|k|} m_T(k;s,t) e^{-|k|(|t|+|s|)} dt ds \right), \nonumber
\end{eqnarray}
where
\begin{equation}
m_T(k;s,t) = \mathbb{E}_{\N_T} \left[ (e^{ik\cdot q_t}-\hat h(k))(e^{-ik\cdot q_s}-\hat h(k)) 
\right].
\end{equation}
We will use two different estimates of this function. One is the uniform bound 
\begin{equation}
|m_T(k;s,t)| \; \leq \; c_1,
\end{equation}
and the other is 
\begin{equation}
|m_T(k;s,t)| \; \leq \; |k|^2 \int_{\R^3} \int_{\R^3} p^T_{s,t} (q_1,q_2) (|q_1| + b)(|q_2| + b) 
d\n^0(q_1)d\n^0(q_2)
\end{equation}
for $|t-s| \geq 1$, where $p^T_{s,t}$ is the probability density with respect to $d\n^0(q_1)
d\n^0(q_2)$ of the measure $\N_T$ with $q_t = q_1$, $q_s = q_2$, and $b, c_1 > 0$ are some 
constants. Using the intrinsic hypercontractivity of the reference measure and the cluster 
expansion of \cite{LM} we have in this case that 
\begin{equation}
|p^T_{s,t} (q_1,q_2)| \; \leq \; c_2
\end{equation}
with some $c_2 > 0$, moreover by the exponential fall-off of the stationary measure of the 
reference process %\cite{BL} 
\begin{equation}
\int_{\R^3} (|q| + b) d\n^0(q) = c_3 < \infty,
\end{equation}
with some $c_3 > 0$. Thus we estimate the exponent in (\ref{expo}) by writing
\begin{eqnarray*}
\lefteqn{
\int_{\R^3} dk \int_{-T}^T \int_{-T}^T \frac{|\hat\rh(k)|^2}{|k|} m_T(k;s,t) e^{-|k| (|t|+|s|)}
dsdt} \nonumber  \\ &&
\leq \int_{\R^3} \frac{|\hat\rh(k)|^2}{|k|} \left( c_1 \int_{- T \leq s,t \leq T \atop |t-s| 
\leq 1} e^{-|k| (|t|+|s|)} dsdt + c_2 c_3^2 |k|^2 \int_{|t|, |s| \leq T \atop |t-s| \geq 1} 
e^{-|k| (|t|+|s|)}
dsdt \right) dk \nonumber  \\ && 
\leq \int_{\R^3} \frac{|\hat\rh(k)|^2}{|k|}\left(\frac{A_1}{|k|} + A_2 \right) dk 
< \infty \nonumber
\end{eqnarray*}
with some constants $A_1,A_2 > 0$. This then yields a non-zero uniform lower bound on 
(\ref{sufi}), and completes the proof of the theorem.
\end{pf}

\section{Hamiltonian in the new representation}
The Hamiltonian $H$ of the infrared regular representation is defined through (\ref{H}). For
having a more concrete expression the natural strategy is to build the Fock space $\cal F$ 
over $L^2({\cal S}'(\R^3), d\g)$ and to transform $H$ unitarily to an operator acting on 
$L^2(\R^3,dq) \otimes {\cal F}$. We do this in three steps. 

First we map $L^2(\R^3,d\n^0)$ to $L^2(\R^3,dq)$ through the similarity transformation 
$R: \phi(q) \mapsto \psi_0(q) \phi(q)$. Thereby the particle Hamiltonian becomes
\begin{equation}
H_{\rm{\tiny{p}}} = -\frac{1}{2}\Delta + V(q).
\end{equation}
Next we deal with the Hamiltonian for the free field. First $\cal U$ defined by (\ref{ufi}) 
is used to map $L^2({\cal S}'(\R^3),d\g)$ into $L^2({\cal S}'(\R^3),d\g_0)$. This map induces
a transformation between the shifted, $\G$, and non-shifted, $\G_0$, Gaussian processes, and 
keeps their generator the same. Then by the Wiener-It\^o transform ${\cal W}: L^2({\cal S}'
(\R^3), d\g^0)$ $\to {\cal F}$ this is further mapped into Fock space (for details see e.g. 
\cite{Oba}). Finally we transform the interaction. By (\ref{H}) in $L^2(\R^3 \times {\cal S}'
(\R^3), dq \times d\g)$ the interaction is multiplication by 
\begin{equation}
\int_{\R^3} \hat\varrho(k) \hat \xi(k) \left (e^{i k\cdot q} -\hat h(k) \right) 
e^{-i k\cdot x} dk
\end{equation}
which under $\cal U$ goes over to the multiplication operator  
\begin{equation}
\int_{\R^3} \hat\varrho(k) (\hat \xi(k) -\hat\gamma(k)) 
\left( e^{i k\cdot q} -\hat h(k) \right) e^{-i k\cdot x} dk
\end{equation}
on $L^2(\R^3 \times {\cal S}'(\R^3), dq \times d\g_0)$. Consider now 
\begin{equation}
{\cal V} = (1 \otimes {\cal W}) (1 \otimes {\cal U}) (R \otimes 1)
\end{equation}
as an isometry from $L^2(\R^3 \times {\cal S}'(\R^3), d\pp^0)$ to $L^2(\R^3,dq) 
\otimes {\cal F}$. Then
\begin{eqnarray}
{\cal V} H {\cal V}^{-1} 
& = & 
\left(-\frac{1}{2}\Delta + V(q)\right) \otimes 1 + 1 \otimes \int_{\R^3} |k| a^*(k) a(k) dk 
\nonumber \\
& \hspace{0.5cm} + &
\int_{\R^3} \frac{\hat\varrho(k)}{\sqrt{2|k|}} 
\left( (e^{i k\cdot q} - \hat h(k)) \otimes a(k) + (e^{-i k\cdot q} - \hat h(k))\otimes a^*(k) 
\right) dk \nonumber \\
& \hspace{0.5cm} - &
\int_{\R^3} \frac{|\hat\rh(k)|^2}{k^2} \hat h(k) \left( e^{i k\cdot q} - \hat h(k) \right) 
\otimes 1 \; dk
\nonumber \\
& \equiv &
H^{\rm{\tiny{ren}}}_{\tiny{\rm{N}}}.
\label{ren}
\end{eqnarray}

The standard Nelson Hamiltonian $H_{\rm{\tiny{N}}}$ corresponds to setting $\hat h \equiv 0$ in
(\ref{ren}). We see that the subtraction of $\hat h$ regularizes the interaction at small $k$ 
and makes $H^{\rm{\tiny{ren}}}_{\rm{\tiny{N}}}$ a well defined Hamiltonian having a ground state
in $L^2(\R^3,dq) \otimes {\cal F}$. In an operator theoretic context, Arai proposed 
$H^{\rm{\tiny{ren}}}_{\rm{\tiny{N}}}$ of the form (\ref{ren}) for the special case $\hat h 
\equiv 1$ and proved that under similar assumptions on the potential as ours 
$H^{\rm{\tiny{ren}}}_{\rm{\tiny{N}}}$ has a ground state \cite{A}. 

Obviously, for obtaining the ground state expectation of an operator $A$ of physical interest, 
one can use $H^{\rm{\tiny{ren}}}_{\rm{\tiny{N}}}$ only in conjunction with the transformation $A 
\mapsto {\cal V} A {\cal V}^{-1}$. Thus expectations of $A$ obtained in the standard way by first 
introducing the cut-off Hamiltonian, taking ground state expectations with it, and subsequently 
removing the cut-off are identical with computing averages of ${\cal V} A {\cal V}^{-1}$ by using 
$H^{\rm{\tiny{ren}}}_{\rm{\tiny{N}}}$. As proved above, the infrared limit is thus correctly taken 
care of. 

\vspace{0.5cm}
{\bf Acknowledgments:} R.A.M. thanks Zentrum Mathematik of Technische Universit\"at M\"unchen 
for warm hospitality and financial support. He also thanks the Russian Fundamental Research 
Foundation (grants 99-01-00284 and 00-01-00271) and CRDF (grant NRM 1-2085) for financial support.


\begin{thebibliography}{99}
\bibitem{A}
Arai, A.: Ground state of the massless Nelson model without infrared cutoff in a non-Fock 
representation, to appear {\it Rev. Math. Phys.} (2001)

\bibitem{AHH}
Arai, A., Hirokawa, M. and Hiroshima, F.: On the absence of eigenvectors of Hamiltonians in a class 
of massless quantum field models without infrared cutoff, {\it J. Funct. Anal.} {\bf 168}, 470-497 
(1999)

\bibitem{F}
Fr\"ohlich, J.: On the infrared problem in a model of scalar electrons and massless scalar 
bosons, {\it Ann. Inst. Henri Poincar\'e} {\bf 19}, 1-103 (1973)

\bibitem{LM}
L\H{o}rinczi, J. and Minlos, R.A.: Gibbs measures for Brownian paths under the effect of an external 
and a small pair potential, to appear in {\it J. Stat. Phys.} (2001)

\bibitem{LMS}
L\H{o}rinczi, J., Minlos, R.A. and Spohn, H.: The infrared behaviour in Nelson's model of a quantum
particle coupled to a massless scalar field, submitted for publication (2001)

\bibitem{Nel}
Nelson, E.: Interaction of nonrelativistic particles with a quantized scalar field, {\it J. Math. 
Phys.} {\bf 5}, 1190-1197 (1964)

\bibitem{Oba}
Obata, N.: {\it White Noise Calculus and Fock Space}, Springer, 1994

\iffalse

\bibitem{AGV}
Arnold, V.I., Gussein-Zahde, S. and Varchenko, A.: {\it Singularities of Differentiable Maps}, 
vol. 2, Nauka, Moscow, 1984 (in Russian)

\bibitem{BFS}
Bach, V., Fr\"ohlich, J. and Sigal, I.M.: Quantum electrodynamics of confined non-relativistic 
particles, {\it Adv. Math.} {\bf 137}, 299-395, 1998

\bibitem{Ger} 
G\'erard, C.: On the existence of ground states for massless Pauli-Fierz Hamiltonians, {\it 
Ann. H. Poincar\'e} {\bf 1}, 443-459, 2000

\bibitem{GLL}
Griesemer, M., Lieb, E.H. and Loss, M.: Ground states in non-relativistic quantum electrodynamics,
preprint, 2000

\bibitem{Nel1}
Nelson, E.: Schr\"odinger particles interacting with a quantized scalar field, in: {\it Proc. Conf. 
on the Theory and Applications of Analysis in Function Space}, W.T. Martin and I. Segal, eds., MIT, 
1964, p. 87

\bibitem{OS}
Osada, H. and Spohn, H.: Gibbs measures relative to Brownian motion, {\it Ann. Probab.} {\bf 27},
1183-1207, 1999

\bibitem{Spo1}
Spohn, H.: Ground state(s) of the spin-boson Hamiltonian, {\it Commun. Math. Phys.} {\bf 123},
277-304, 1989

\bibitem{Spo2}
Spohn, H.: Ground state of a quantum particle coupled to a scalar Bose field, {\it Lett. Math. Phys.} 
{\bf 44}, 9-16, 1998
\fi 
\end{thebibliography}
\end{document}